# DYNAMIC APERTURE AND SPACE CHARGE EFFECT STUDIES FOR THE RECYCLER RING FOR PROJECT-X[*]

M. Xiao[*], L.G. Vorobiev and D.E. Johnson,
Fermilab, Batavia, IL 605010, U.S.A.

*Abstract*

A simplified Recycler lattice was created to fine tune injection straight, ring tune, and phase trombone. In this paper, we will present detailed modifications for further optimization of Recycler lattice which requires the investigation of tune footprint and dynamic aperture based on higher order momentum components of the magnetic fields, together with the space charge effects.

## INTRODUCTION

Project X [1] is a multi-MW intense proton source that provides beam for various physics programs. The Recycler ring will be used as a proton accumulator where H$^-$ would be injected and converted to protons. Protons are provided to the Main Injector and accelerated to desired energy. The injection system for converting H$^-$ to protons in Recycler is a multi turn stripping system, see Fig. 1. A simplified toy lattice was created to fine tune the injection insertion, ring tunes and phase trombone for the Recycler ring [2]. In this paper, a realistic lattice was created by using the measured magnetic field for all the magnets and further optimization of this lattice was completed. Based on this lattice, the tune footprint and dynamic apertures in the present of higher order multipole components of the magnetic fields have been investigated and are presented. Space charge effect is another issue for this lattice since the beam intensity at the end of injection reach 1.6E14, which is 2 order of magnitude larger than the existing beam intensity in the Recycler ring. The preliminary results of the space charge effect study is also presented in this paper.

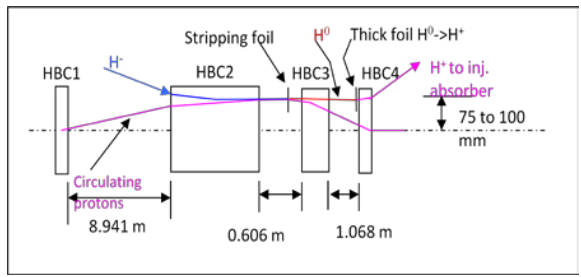

Figure 1: injection insertion.

## REAL RECYCLER LATTICE FOR PROJECT X

To accommodate the injection system in the Recycler ring, a 21.5 m long drift space is designed by converting the existing FODO lattice in RR10 straight section into a doublet, shown in Fig. 2. Instead of the standard ideal magnetic field used in the toy lattice, the measured magnetic fields, up to 8$^{th}$ order multipole components, have been implemented in the real lattice for Project X. In addition, the RR30 straight section was converted to a FODO lattice with standard permanent quads, and the trim quads in RR60 phase trombone straight section are set to zero. To get nominal tunes (25.425, 24.415),

- the end-shim field of each gradient magnet in the arc cell were adjusted, so that the phase advances of the arc cell changed from $\mu_x$ =83.624$^o$, $\mu_y$ =78.290$^o$ to $\mu_x$ =85.236$^o$, $\mu_y$ =79.007$^o$
- added additional trim quads in the dispersion suppressor sections on either side of the RR10 to match the two ends of the RR10 injection insertion to the whole ring

The lattice was shown in Fig. 3. The chromaticities in the Recycler ring were designed to be corrected by body sextupole components and the sextupole components of the end-shims of each dipole gradient magnets. The

Work supported by Fermi Research Alliance, LLC under Contract No. DE-AC02-07CH11359 with the U.S. Department of Energy

chromaticities are now (-1,-1) with the measured magnetic field. For additional chromaticity corrections, there are 8 and 16 sextupoles in horizontal plane and vertical plane respectively. They are set to 0 for this lattice.

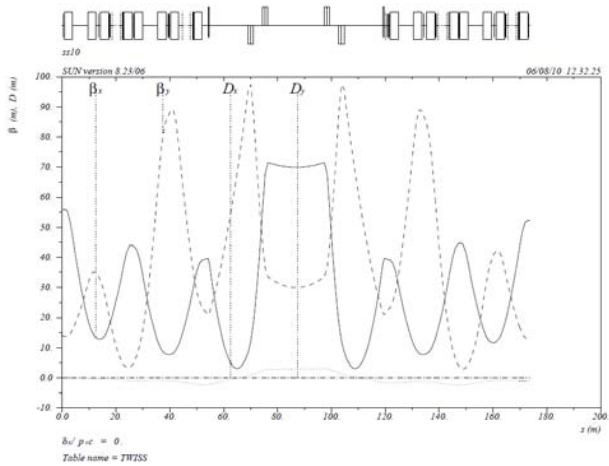

Figure 2: The lattice in RR10 with symmetric structure for injection

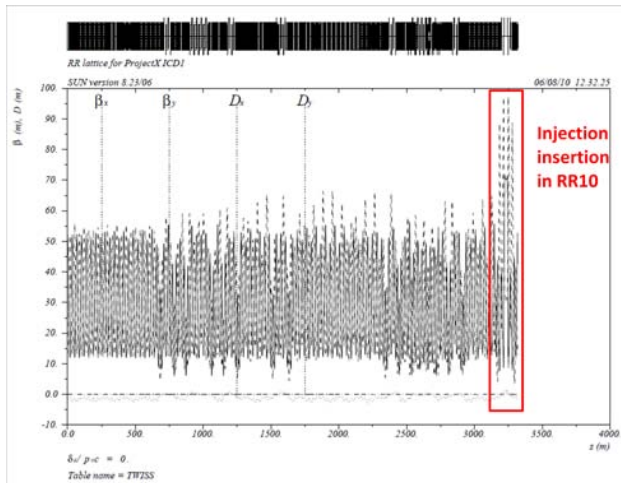

Figure 3: Real Recycler lattice for project X. Included are measured magnetic field up to 12[th] order of multipole components, nominal tunes are (25.425,24.415)

## DYNAMIC APERTURE STUDY

The scenario of beam injection into the Recycler ring for Project X is shown in Fig. 4. The beam of 1.6E14 is divided into 6 injections with the time interval of ~100 ms, total time of the particles circulating in the Recycler is about 0.5 second, which is about 45,000 turns (Recycler Revolution period is 11.12 ms). The 95% normalized emittances of 25 $\pi mm.mrad$ in both planes are achieved after painting.

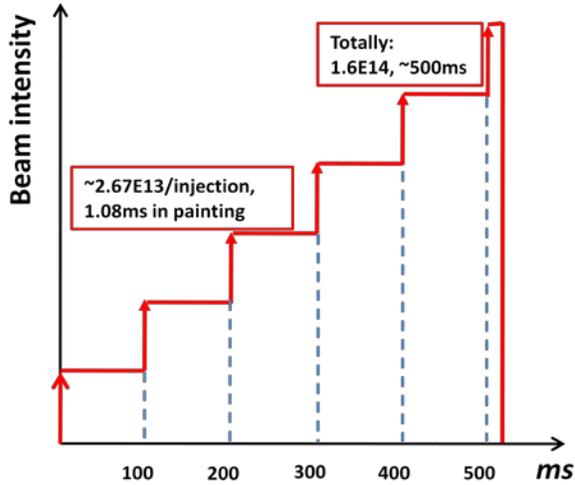

Figure 4: The scenario of the Beam injection into the Recycler ring

The dynamic aperture tracking was done using MAD (*Ver*8.23). Particles are launched with a distribution of amplitudes with neighbouring particles differed in amplitude by either $1\sigma_{xo}$ or $1\sigma_{yo}$ which are the beam sizes at launch point. For each fixed *x*-amplitude, we search the largest *y*-amplitude for which the particles survive 45,000 turns. We also check that particles with smaller y amplitudes are stable over this number of turns. This is repeated for several *x* amplitudes until the largest *y*-amplitude falls to 0. The dynamic aperture is then defined as the average of all the largest stable radial amplitudes.

The gradient dipole magnet in arc cell is 4.496 meters and 3.099 meters in dispersion suppressor cell. Previous experiences [3] show that the magnet needs to be sliced into at least 16 pieces, each with $(1/16)^{th}$ of the integrated strength of the whole single kick, for the non-linear lattice model to incorporate the multipole kicks. We observed that the beta functions are not varying rapidly along the length, but the phase advance change of ~7 degree along the length of each gradient magnet. When there are several kicks along the length, each of these occurs at a slightly different phase, the resultant of these somewhat incoherent kicks will always be smaller than a single coherent kick which has the same length as the sum of all the individual kicks. This is a qualitative explanation of the effects of several incoherent kicks. For this tracking, each gradient magnet is sliced into 16 pieces. Fig. 4 gives the dynamic aperture of the Recycler ring for Project X. Particles with 3 constant momentum deviations of $\Delta p/p=0$, $\Delta p/p=0.2\%$ and $\Delta p/p=-0.2\%$ are tracked. The average dynamic aperture is $10\sigma$ in radial for the particles with $\Delta p/p=0$, and $8.4\sigma$ in radial for the particles with $\Delta p/p=0.2\%$ or $\Delta p/p=-0.2\%$.

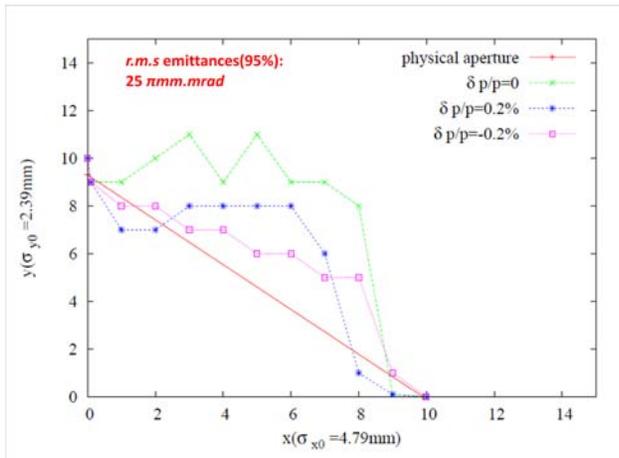

Figure 5: Dynamic aperture in the Recycler ring for Project X after 45,000 turns for 3 cases of constant momentum deviations: $\Delta p/p=0$ (Green), $\Delta p/p=0.2\%$ (Blue) and $\Delta p/p=-0.2\%$ (magenta). Also shown is the physical aperture (Red) of the Recycler beam tube. The dynamic aperture of the particles with $\Delta p/p=0$ exceeds the physical aperture.

The physical aperture was calculated from the Recycler beam tube, size of 47.625 mm×22.225 mm. Fig. 4 shows that the dynamic aperture of the particles with Δp/p=0 after 45,000 turns exceeds the physical aperture. Actually, the emittance only reaches 25 π mm.mrad after the last injection and only last for less than 1,000 turns. Therefore it would be still safe for the off momentum particles circulating for 45,000 turns.

Tunes for the particles with up to $6\sigma_{xo}$ or $6\sigma_{yo}$ amplitudes are calculated by tracking the particle 1024 turns, and perform a FFT from the calculated turn by turn data. They are plotted in the tune diagram shown in Figure 5. The resonance lines are $7^{th}$, $9^{th}$ and $12^{th}$ order. Most of the tunes are lined within $12^{th}$ resonances lines. Our nominal tunes are also on the $12^{th}$ resonance lines, which should be safe since the Recycler ring now is a proton accumulator, driving strength of higher order resonances is not large enough to drive the particle out in a short time(~0.5 second).

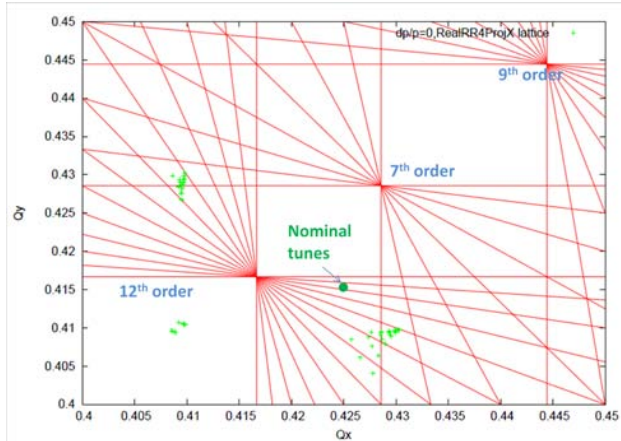

Figure 6: Tune footprint of the particles with $6\sigma_{xo}$ or $6\sigma_{yo}$ amplitudes. Also shown is the nominal tunes (green solid circle).

## SPACE CHARGE EFFECT STUDY

To reach the level of 1.6E14 protons, circulating in the Recycler in Project-X, one needs a multi-turn H$^-$ stripping injection, using a painting procedure. The detailed injection painting was considered in [4]. In our numerical studies we skip the injection cycle, have been taking into consideration the resulting beam of 1.6E14 protons and simulated the dynamics during 800 turns using the multi-particle code ORBIT [5].

ORBIT package was extensively used for SNS design and operations and combines the tracking with the space charge physics. It allows easy extensions, by adding new modules to the existing library of C++ classes. Our current version [6] was equipped with a corrected $2^{nd}$ order transfer matrices, a foil hits module and the procedure for injection painting of the longitudinal train of arbitrary chopped microbunches

The calculations were done with 50,000 macro-particles. The initial phase space distribution was assumed to be a bi-Gaussian with *95%* normalized emittamce of 25 *πmm.mrad*. Our first step is to check how the space charge affects the particles with the small amplitudes. So far a linear toy lattice for the Recycler was implemented and the Twiss and Transfer matrices from MAD were imported into ORBIT. The space charge was taken into account with full generality. We would include the lattice non-linearity into the model later.

The evolution of the space charge tune spread after turn1, turn 403 and 800 turns are shown in Fig.7 (a) (b) and (c)t respectively. We can see most of the particles (30,000 among of 50,000) are in the core of the beam distribution (represented by the red colour) after turn 1, but they smear out (10,000 left after turn 803), and the tunes are spread and shifted down by 0.01 in both planes

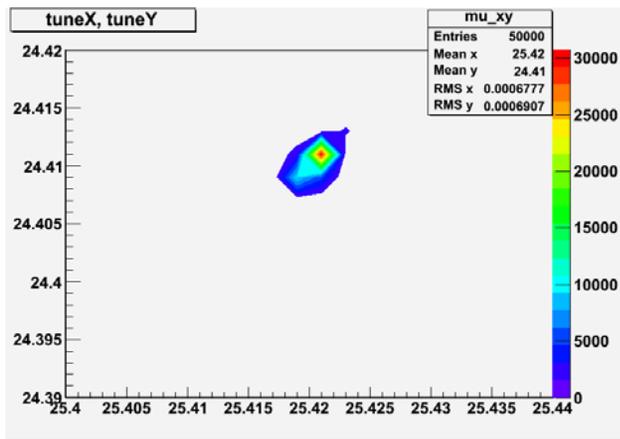

(a) Tune spread after Turn1

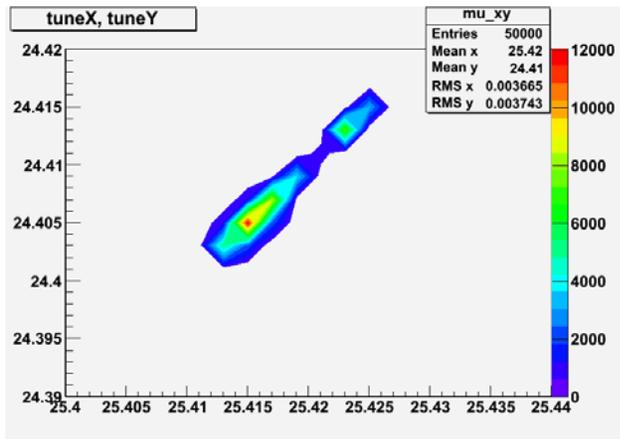

(b) Tune spread after Turn403

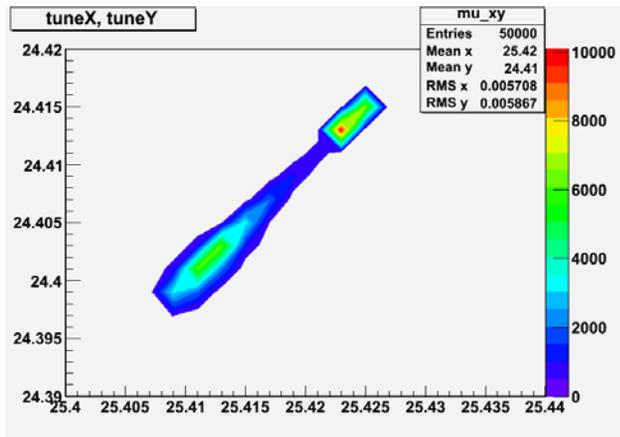

(c) Tune spread after turn 800

Figure 7: the evolution of the space charge tune spread.

We place space charge tune spread on tune diagram for turn 2 and turn 803, shown in Fig. 8 We can see after 803 turns, the space charge effect tunes are spread to $5^{th}$ order resonance lines, which would potentially cause emittance growth.

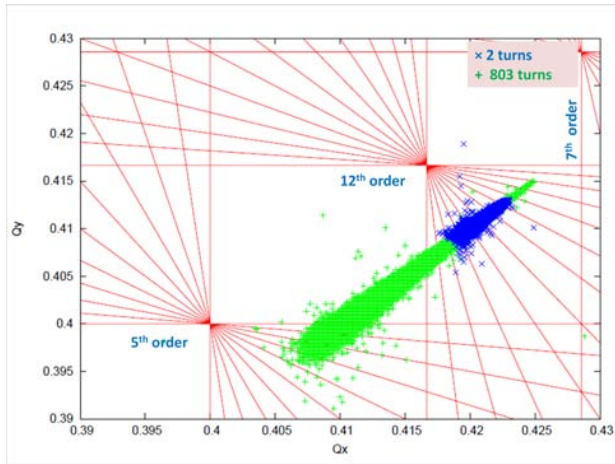

Figure 8: Space charge tune footprint

Fig. 9 presents the beam distribution in (*x,x'*) and (*y,y'*) planes after 2 turns and after 803 turns. We can see from this plot the growth rate is about 2-3%

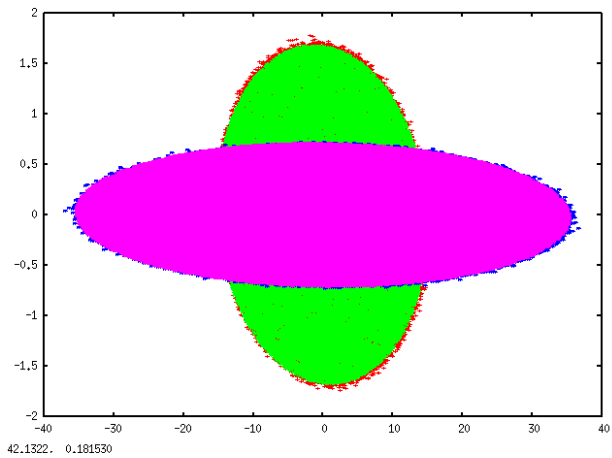

Figure 9: Beam distributions in phase space. (x,x'):Red – after 800 tuns, Green – after 1 turn; (y,y'): Blue – after 1000 turns, Magenta – after 1 turn.

## CONCLUSION

Dynamic aperture was found to be $10\sigma$ in radial for the beam with the *95% normalized* emittances of 25 $\pi mm \cdot mrad$ in the Recycler lattice for Project X. This was obtained for the measured magnetic field for all the magnets up to 8[th] order momentum components. It is larger than the physical aperture in the Recycler ring. Space charge effect study show that the emittance growth for the corn of the beam is about 2-3%. The space charge tune spread down to the 5[th] order resonance lines could potentially lead to emittance growth or beam lifetime issues. Shifting the base tune up by .005 in both planes should elevate this issue.

Further simulation study with more particles and longer turn tracking, together with magnetic field errors will be done soon. Also sfter the next shutdown we should look at the strength of the 5[th] order resonance with protons.